# Neural Digital Twins: Toward Next-Generation Brain-Computer Interfaces

Mohammad Mahdi Habibi Bina[1], Sepideh Baghernezhad[1], Mohammad Reza Daliri[1*], Mohammad Hassan Moradi[2*]

***Abstract*— Current neural interface such as brain-computer interfaces (BCIs) face several fundamental challenges, including frequent recalibration due to neuroplasticity and session-to-session variability, real-time processing latency, limited personalization and generalization across subjects, hardware constraints, surgical risks in invasive systems, and cognitive burden in patients with neurological impairments. These limitations significantly affect the accuracy, stability, and long-term usability of BCIs. This article introduces the concept of the Neural Digital Twin (NDT) as an advanced solution to overcome these barriers. NDT represents a dynamic, personalized computational model of the brain–BCI system that is continuously updated with real-time neural data, enabling prediction of brain states, optimization of control commands, and adaptive tuning of decoding algorithms. The design of NDT draws inspiration from the application of Digital Twin technology in advanced industries such as aerospace and autonomous vehicles, and leverages recent advances in artificial intelligence and neuroscience data acquisition technologies. In this work, discussed the structure and implementation of NDT and explore its potential applications in next-generation BCIs and neural decoding, highlighting its ability to enhance precision, robustness, and individualized control in neurotechnology.**

***Index Terms*— Brain-Computer Interfac, BCI, Digital Twin, Deep Learning, Neural Decoding***

## I. Introduction

Our understanding of the human brain has been transformed during the last thirty years. It is now seen as a modelable information network. Brain Computer Interfaces (BCIs) take advantage of this transformation [1], [2]. They bypass the neuromuscular pathways to interpret commands directly as translated from neural activity. Although this technology is a valuable and essential resource to patients suffering from locked-in syndrome and cognitive modeling [3], it is unreliable. Persistent challenges—specifically signal instability and poor cross-subject generalization—expose a harsh truth: our current models still struggle to capture the brain's full complexity [4]–[5].

In this effort, computational neuroscience is essential for elucidating the principles that regulate cerebral information processing [6]. This discipline seeks to elucidate the intricate mechanisms of neuronal interactions, the spatiotemporal dynamics of cerebral activity, and their mechanistic correlations with behavior and cognition through mathematical modeling, stringent signal analysis, and neural network simulations [6]–[7].However, the inherent complexity of the brain, the strong dependence of neuronal activity on the instantaneous perceptual and behavioral state, and the difficulty of multifaceted and stable recording of neural signals have led many models to be limited representations of reality and to perform less than expected in natural environments [7]–[8].

Therefore, where these two domains meet, that is, the point of connection between brain signals, behavior, environment, and an individual's internal state, the need for a more comprehensive approach is felt more than ever [8]. The combination of the brain's instantaneous dynamics, human behavioral patterns, and tissue physiological characteristics presents challenges, such as long-term instability, a lack of coordination between multi-source data, and the absence of reliable predictive mechanisms [4], [8]. Consequently, even the most advanced BCI systems face fundamental limitations when confronted with the natural and variable realities of the human brain [3].

To overcome these barriers, we must look beyond biological discovery alone. As the timeline suggests in Fig. 1, progress here stems from a convergence of "hardware push" and "software pull" rather than a simple linear evolution. Early research was shackled by computational scarcity, forcing scientists to rely on reductionist methods like the classic Hodgkin and Huxley model (1952) [4]. Although game changed definitively with the arrival of GPU computing in 1999 [5].

In this case, a "hardware push" has previously been described as the removal of computational limitations and the provision of the parallel processing capabilities required to simulate a sophisticated, interdependent, and complex system, rather than a simple chained model of a unit or system of biological units.

At the same time, a shift in the understanding of complex neural computation from a "software pull" has been described. The field had been decoding individual units (neurons or single neural units) colocated in neural circuits, but this changed to the analysis of neural of ensembles or populations. The spark for this was Georgopoulos 1982 [6]; the revolution occurred in the work of Krishna Shenoy on neural manifolds. Shenoy showed how the activations from a population of neurons formed the surface of a low-dimensional manifold and that this population of neurons was a dynamical system [12]. The field of complex system and dynamical systems theory requires the tools to model and to analyze such systems. This groundwork has and continues to employ deep learning techniques the POYO model is an example of this where neural spikes (tokens) are input into a transformer system[7] where and the model [8].

[1] Neuroscience & Neuroengineering Research Laboratory, Biomedical Engineering Department, School of Electrical Engineering, Iran University of Science and Technology (IUST), Tehran, Iran

[2] Department of Electrical Engineering, Faculty of Engineering, Bu-Ali Sina University, Hamedan, Iran

*Corresponding authors: daliri@iust.ac.ir ; mhmoradi@basu.ac.ir

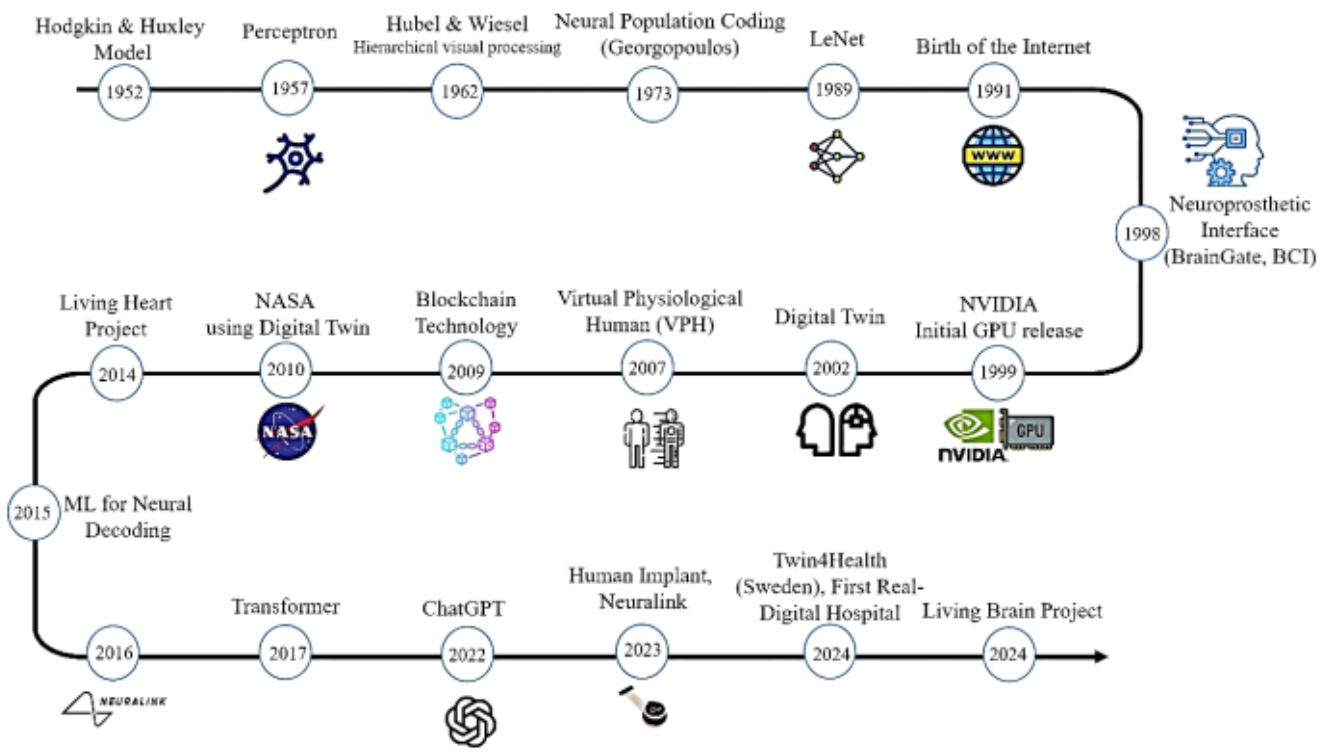

Fig. 1. Timeline illustrating the major developments in brain–computer interface research from 1952 to 2025.

In this context, the concept of the Digital Twin creates a new frontier in neuroscience and BCI. It is not intended as an instant remedy, but rather as a foundational framework enabling highly accurate, time-variant, and personalized representations of neural dynamics [15]. This method makes it easier to deal with and analyze previous issues by systematically integrating neural, behavioral, environmental, and physiological data into a model that is always changing [16]. This is because, for the first time, a system is being formed that is capable of reflecting an individual's real-time status.

In the past decade, digital twins in neuroscience have moved beyond a theoretical concept and become a practical tool for deeper brain modeling [9]–[10]. In this view, a Digital Twin is not just a computational model; it is a dynamic, data-driven version that is precisely tailored to an individual's neurobehavioral characteristics and has a two-way interaction with the physical world [10]–[12]. The combination of multimodal data, from intracortical signals to non-invasive signals to behavior, cognitive state, and physiological indicators, makes it possible to create a digital, real-time synchronized version that can predict neural patterns, identify instantaneous changes, and optimize the settings of neuro-computational systems in a personalized manner [13]–[15].

BCI, computational neuroscience, and Digital Twins' convergence shape the architecture of upcoming neuro-intelligent systems[16]. In this case, the brain is not just a data producer but also a valuable contributor. The system collaborates with its digital twin to incorporate neurophysiological data and deep learning to enhance understanding of the neural data calibrated, behavior and the surroundings. This brings a situation-specific responsiveness to people with motor or cognitive challenges who acquire a system that actively modifies to their changing neural patterns [17]–[20]. The convergence of these fields removes the hard line between computation and biology and enables the construction of advanced personalized neural models, altering the approaches to designing and evaluating brain-driven systems.

The rest of the paper is organized in the following order: in section II, describes the data sources and search strategy employed in this study, detailing the criteria and procedures used for literature selection. Section III presents the theoretical background, covering the fundamentals of DT concepts, including DT levels and types, architectural frameworks, and technological aspects spanning industrial cyber–physical systems to neural dynamics, along with an overview of BCIs. Section IV serves as the core of this study: it first identifies the major challenges in current BCI systems and then introduces the DT paradigm as a transformative solution, culminating in the proposed BCI–DT framework. Section V explores the emerging applications of Digital Brain Twins in neuroscience. Finally, Section VI concludes the paper and discusses future research directions.

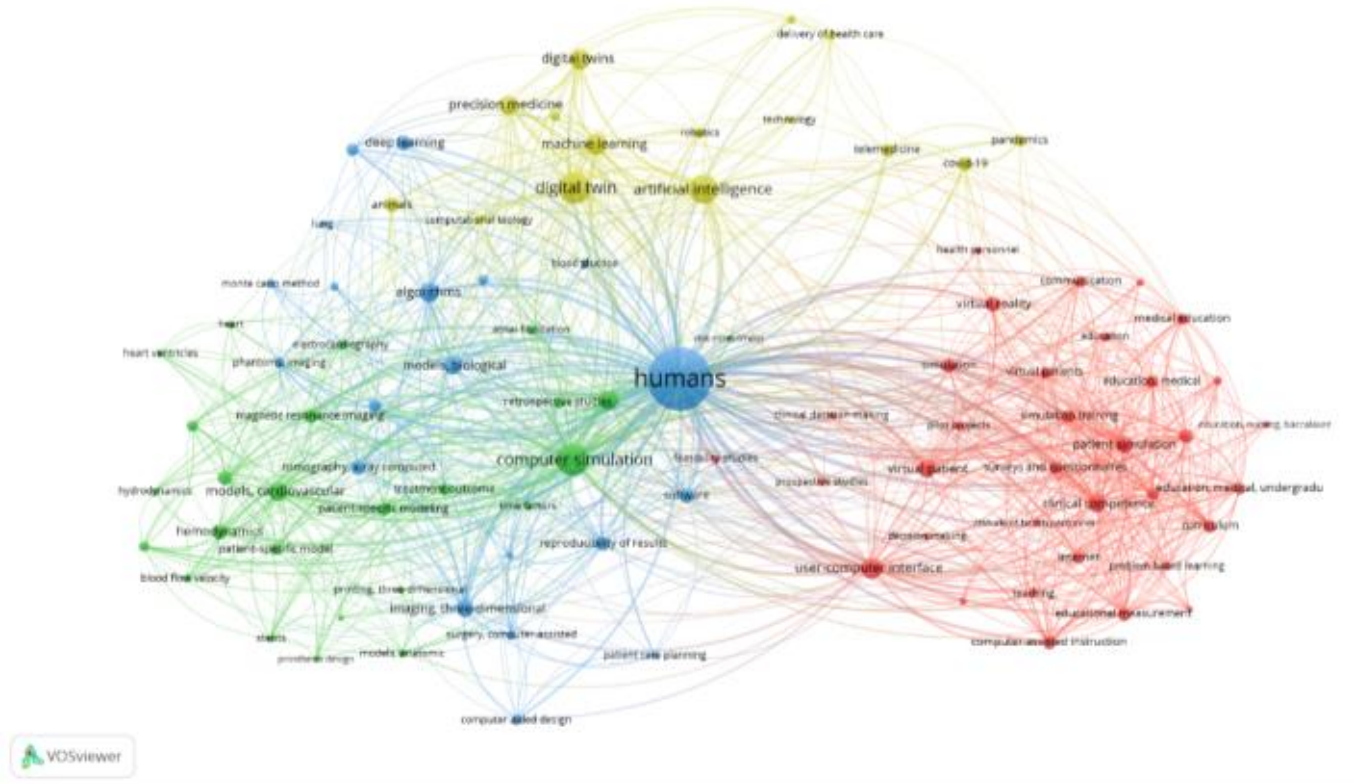

Fig. 2. Keyword frequency map generated from PubMed records and visualized with VOSviewer

## II. Data Source And Search Strategy

As illustrated in Fig. 3, this paper's review employs the PRISMA (Preferred Reporting Items for Systematic Reviews and Meta-Analyses) technique to select studies and reduce the search space [19].To construct a robust and unbiased search query, a data-driven pre-search refinement stage was conducted before the formal literature search. A thorough collection of potential keywords from foundational texts in BCI, neural signal processing, computational neuroscience, and digital twin frameworks was initially compiled. To reduce personal bias in keyword selection, metadata from initial exploratory searches—including titles, abstracts, and author keywords—were extracted and analyzed with VOSviewer [20]. This bibliometric analysis visualized keyword co-occurrence networks (see Fig. 2), where node size corresponds to frequency of occurrence and clustering indicates conceptual relationships. This iterative process improved the search string so that it could effectively find digital twin intersections, computational modelling, and neural engineering.

Database selection and technical limitations: A formal search was conducted in PubMed and Scopus, which were chosen due to their extensive coverage of the biomedical engineering and neuroscience literature. Certain databases were excluded due to methodological and technical constraints. ScienceDirect and IEEE Xplore were specifically excluded due to their search interfaces, which impose stringent limitations on the number of Boolean operators and keyword entries. Because of this, these platforms couldn't handle the complexity and length of the optimized search string without making the query less specific in a big way.Since most high-impact engineering journals, like IEEE Transactions, are indexed in Scopus, this exclusion doesn't mean a big loss of coverage. The ASCE database was removed to maintain the accuracy of review, citing concerns about the reproducibility of its search algorithms.

## A. Methodical Approach to Literature Review

### *1)* Identification

On November 8, 2025, two authors (Habibi and Baghernejad) independently used the finalized keyword set without MeSH term constraints to perform the advanced searches [85]. Two main conceptual clusters were intersected by the search approach using the Boolean AND logic:
1.1. Terms associated with digital twins (such as virtual counterparts, twinning, and digital twins).
1.2. Terms pertaining to neuroscience and BCI (e.g., neural decoding, brain-computer interface, EEG).

All lexical variations were covered using wildcards (e.g., twin*). A total of 606 records were found in the first search (338 in Scopus and 218 in PubMed). After duplicate entries were eliminated, 508 distinct records remained for screening.

### *2) screening*

Title/abstract screening and full-text evaluation were the two successive processes that comprised the screening phase. Records that did not satisfy the following requirements were disqualified during the first screening.
2.1. Relevance to Digital Twins: Research centered on non-biological engineering twins (e.g., industrial turbines) or employed the term "twin" metaphorically was excluded.
2.2. Relevance to Neural Interfaces: Articles that addressed physiological systems unrelated to the central nervous system, such as models of the heart or musculoskeletal systems, were not included.
2.3. Type of Publication: Evaluation articles, editorials, viewpoints, and commentary were omitted to guarantee that the evaluation benchmarks methodological advancements based on primary empirical data. This exclusion criterion was essential to ensure that the results were based on primary computational models and experimental frameworks rather than secondary summaries. Consequently, 179 articles qualified for full-text examination after 329 article records were eliminated.

### *3)* Eligibility and Inclusion

We thoroughly reviewed the other 179 full-text papers to make sure they met the set standards.
3.1. Studies without an original computational model were omitted.
3.2. BCI applications or brain signals (such as EEG or fMRI and etc) were not used.
3.3. They briefly discussed digital twins as a concept that lacks methodological integration.
Finally, 23 papers were included in the final qualitative synthesis and analysis as they met all the inclusion requirements.

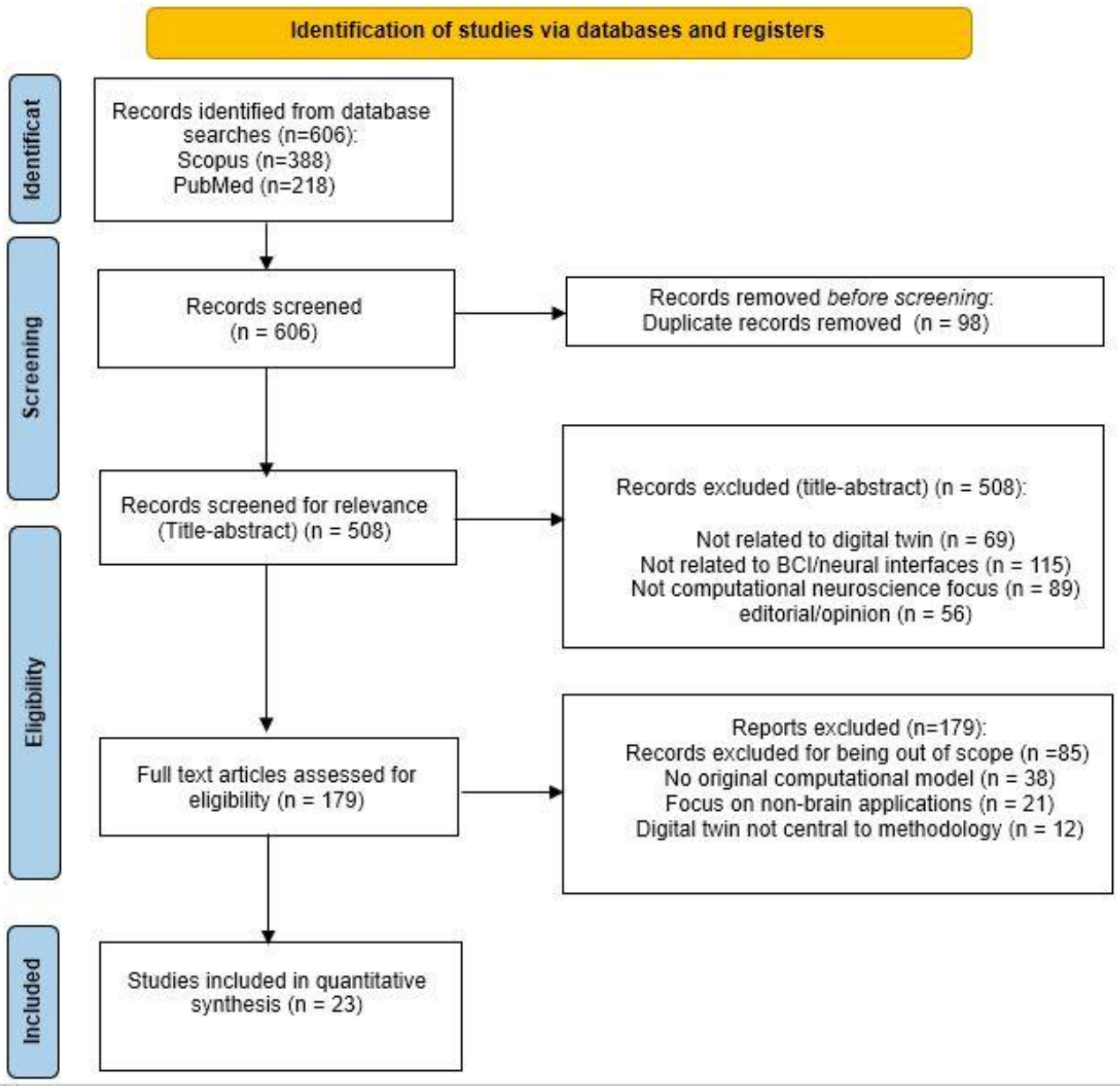

Fig. 3. PRISMA flow chart of literature search procedures.

# III. Theoretical Background

## A. Fundamentals of Digital Twin (DT)

A Digital Twin is a dynamic, interactive, and synchronized digital representation of a living or non-living system, object, or process that simulates, predicts, and optimizes its behavior thru a two-way, real-time relationship [21], [22]–[23]. This technology processes real-time data, like sensor data from the Internet of Things (IoT) [24], [25], EEG signals in neuroscience [26], or LiDAR data in self-driving cars [27] and combines it with scientific models, like physical equations, neural dynamics, or AI and machine learning algorithm. Through data assimilation, techniques such as the Kalman filter and ML models integrate real-world inputs into the digital framework, thereby correcting the model's trajectory to reflect reality[28], [23]. Therefore, the digital twin evolves over time with the physical system [23].

The most significant feature about a digital twin is that it can interact with the real system .This means that changes to the digital model can lead to changes or improvements in the real system, and changes to the real system are automatically shown in the digital model.This characteristic makes the digital twin highly valuable in tackling issues regarding BCIs [70], [74].

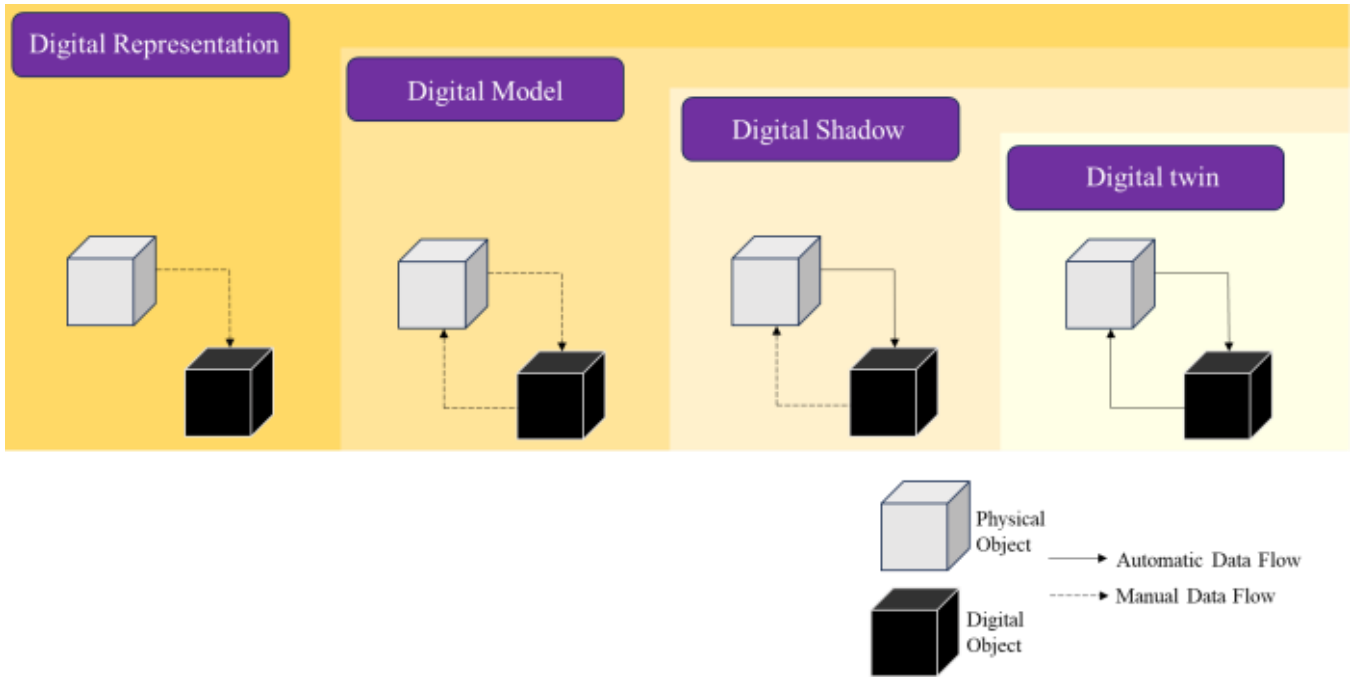

Fig. 4. Visualization of the four-level digital hierarchy: Digital Representation, Digital Model, Digital Shadow, and Digital Twin.

## B. Levels of Digital Twin and Their Types

In various studies, the digital twin has been defined across different dimensions and levels [29], [30]. Four main levels can be categorized as follows, as illustrated in Fig. 4.

### 1) *Digital Representation*

This is the lowest level, and at this level, there is a digital replica of a physical system and a virtual model is created. This virtual model is not real time on line with the physical system, and therefore is limited to visualization, simulation, and theoretical analyses .

### 2) *Digital Model*

In a digital model there is a more detailed, sophisticated, and static replica of the physical system that has more advanced structural, geometric, and functional attributes. Unlike a digital representation, a digital model has the potential for more advanced analysis and simulation of system behaviors. However, it does not contemporaneously interface with the real system.

### 3) *Digital Shadow*

The functional digital twin, also known as the digital shadow, is a digital model with the capability to receive real-time data from its physical counterpart. This level provides a dynamic representation of system behavior, where information flows one-way from the real system to the digital model. Applications of this level include medical device design, surgical planning, in silico clinical trial design, and randomized controlled simulations. The digital shadow can be used for performance prediction, process optimization.

### 4) *Digital Twin*

The most advanced level is the intelligent digital twin, which not only possesses the capability for real-time data acquisition and model updating but also incorporates artificial intelligence algorithms. This twin is capable of learning, reasoning, decision-making, and autonomous action. It can communicate with other digital twins and systems and can perform complex multi-stage processes, including predictive analytics and automated optimization.

## C. Digital Twin Architecture

Digital Twin (DT) approach is no longer just an idea. It has become an integral part of dynamic simulation and intelligent systems. More than just a digital replica, a DT is an adaptive entity that evolves alongside the physical counterpart in real time. A DT, on the other hand, is a living system that constantly takes in data from the real world to help it make smart decisions. The system is built on three layers that work together: Physical Layer, Virtual Layer, and Digital Thread that connects them all[21], [31].

### 1) *Physical Layer*

This layer represents the real-world entity in the physical world; the primary source of data whose behavior needs to be modeled or controlled. This entity can range from an industrial machine or aircraft to a biological process such as a body organ or nervous system. In the field of neuroscience and BCI, this layer can include the brain, neural signals and physical equipment connected to it, such as neural prostheses[21], [32], [33].

### 2) *Virtual Layer*

This section is the digital twin or advanced software model of the physical system that operates in computational space .Using static data, along with machine learning and system models, the virtual tier predicts and reconstructs the physical entity's behavior.For example, in a neural system, a digital twin can estimate an individual's motor intent using brain signal data and predict the desired output for controlling a prosthesis. This layer, in addition to simulation, also has the ability to self-optimize and adaptively learn from new data[21], [32]–[33].

### 3) *Communication Layer or Digital Thread*

This layer acts as a neural network connecting the physical and digital worlds, ensuring data flow in both directions [82].Sensory data is collected from physical entities thru sensors, such as IoT sensors or neural electrodes, and sent to a digital model; the analysis results and predicted decisions are then returned from the digital model to the physical system to establish real-time feedback and control . This data exchange is carried out continuously through data assimilation and sensor fusion techniques, ensuring the accuracy and stability of the system [21], [32]–[33].

The merging of these layers gives way to a closed, dynamic loop, in which data is passed from the real world to the virtual world, while insights are passed back to the real world .This two-way process means that the Digital Twin is more than a passive imitation; it is a mechanism that improves performance and provides predictions on how things should perform in the future [32], [34] .Fundamentally, we view the Digital Twin not merely as a static mechanism, but as a self-evolving ecosystem that bridges the physical-digital divide to enable adaptive control .In complex domains like Neuroengineering and BCI, this framework promises to redefine how we predict, govern, and enhance neural decoding systems.

## D. Technological Aspects of Digital Twins: From Industrial Cyber-Physical Systems to Neural Dynamics

This section investigates state-of-the-art DT-based methods across six key domains: aerospace, smart energy, cognitive vehicles, smart manufacturing, health, and neuroscience. As shown in Fig. 5, while these fields differ significantly in practice, they share a common technical goal: moving from static models to dynamic systems that evolve in real-time [28].

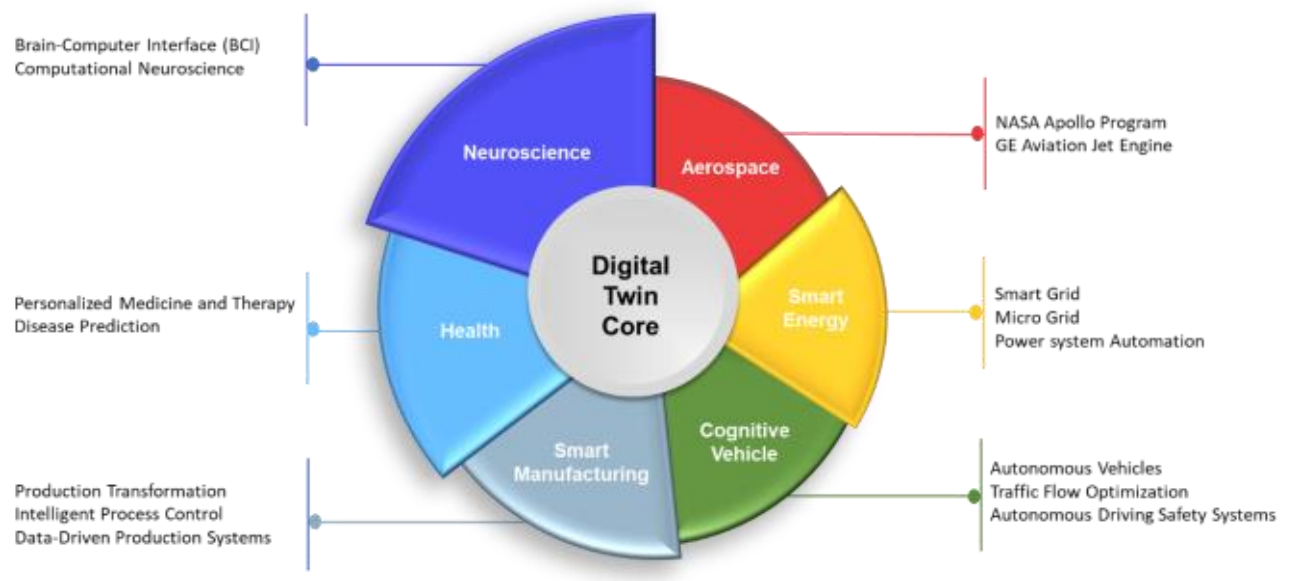

Fig. 5. Central Digital Twin Core surrounded by six key application domains: aerospace, smart energy, cognitive vehicles, smart manufacturing, health, and neuroscience.

A major shift since 2019 involves moving from static engineering models to dynamic, data-centric platforms [31]. These systems now integrate IoT, cloud computing, and machine learning pipelines to create a continuous feedback loop. Crucially, the virtual model no longer just mirrors the physical world; it actively informs and optimizes it [32].

1. Aerospace

The idea of a digital twin came from NASA's "live models" used during the Apollo 13 crisis in the 1960s [23], [35]–[38]. Engineers used ground-based simulators to test control sequences before sending them to the damaged spacecraft to make sure the crew got home safely [35]. These systems didn't have modern real-time data connectivity, but they did show how to mirror physical assets [39]. This approach evolved through Gelernter's "Mirror Worlds" [40] and was later formalized by NASA's John Vickers in the early 2000s [41]. Vickers defined the framework that we recognize today: virtual models continuously updated by mission data[42] [38].

In contemporary aviation, the focus has expanded to cover routine fleet management. Modern aircraft generate massive datasets via IoT sensors, which machine learning algorithms then analyze.The two together lets operators see structural problems ahead of time and plan maintenance schedules in advance.

2. Smart Energy

In smart energy systems, digital twins have revolutionized energy management. They enable precise monitoring, help predict changes in supply and demand, and improve operational strategies[43][44]. In microgrids, digital twins combine data from IoT sensors and measurement devices to model and predict short-term changes in load and renewable energy generation. This feature enables the implementation of predictive control policies for storage systems, demand response programs, and the coordination of distributed resources[45]. This improves voltage and frequency stability while reducing operational losses [46]–[47].

A digital twin is a live, data-driven model of the transmission and distribution network at the level of smart grids and larger power systems. This model constantly reconstructs the network topology, equipment status, distributed generation, and consumer profiles in near real-time [48]–[49].This virtual model allows operators to test various fault scenarios, large-scale integration of renewable energy, dynamic pricing mechanisms, and grid recovery strategies. Before implementing changes to the physical network, their impact on stability, reliability, and cost can be evaluated [50]–[51]. Since 2019, this approach has gained attention in research related to microgrid management, flexible markets, and smart island strategies[52].

3. Cognitive Vehicle

In intelligent transportation, a digital twin provides a real-time 3D representation of a vehicle, its internal subsystems, and its surroundings, including road conditions, traffic, and infrastructure. These twins are directly connected to internal sensor data (LiDAR, radar, cameras, GPS, and IMU) and infrastructure-based data sources [53]–[57]. These settings provide a safe virtual space to test high-risk, rare, or expensive scenarios such as accidents, extreme weather, unmarked intersections, and interactions with other factors. As a result, it has become possible to train and evaluate autonomous driving algorithms on a large scale without exposing real vehicles to risk[58]–[59]. From a cognitive perspective, a digital twin can use predictive models and filtering techniques, such as the Kalman filter, to predict the behavior (speed, path, and maneuvers) of an autonomous vehicle and how it interacts with traffic and responds to noisy sensor inputs. This facilitates decision-making regarding routing, safety, and energy consumption[60]–[61].Automakers like Tesla use a nearly digital twin architecture connected to their fleet. This allows them to monitor mechanical and behavioral issues in real-time and send software updates wirelessly. When vehicles are in use, they become living laboratories that provide continuous feedback for improving virtual models and self-driving algorithms[62].

4. Smart Manufacturing

In the field of smart manufacturing, digital twins have established themselves as the cornerstone of digital transformation, seamlessly expanding from discrete machinery and production lines to comprehensive facility ecosystems and supply chains. Its application is most evident in areas such as predictive health management, spatial layout optimization, minimizing operational downtime, and conducting precise scenario simulations for capacity scaling or asset reconfiguration[21], [63] .These capabilities are based on the continuous collection of data from the convergence of operational and information sources, including field sensors, PLCs, MESs, and ERP architectures [64]. Additionally, in the field of domestic logistics and industrial environments, strategic collaborations – exemplified by the partnership between NavVis and AWS – demonstrate how advanced 3D scanning, machine vision, and real-time object recognition can be utilized to create high-quality digital twins of structural assets and warehouse facilities, enabling automated and optimized workflows for item routing, storage, and retrieval[65]–[67]. Research by Jaghotia and colleagues shows that robots equipped with digital twins and machine vision capabilities can eliminate manual placement tasks, reduce human error, and significantly increase operational efficiency in industrial pick-and-place processes[65].

5. Health

In healthcare, a digital twin is a changing virtual model of a patient, a clinical workflow, or even an entire hospital environment. This computational model can track a patient's physiological state in real time and help with predictive analysis by combining data from different sources, such as electronic health records (EHRs), physiological signals, and high-resolution data from wearable sensors [31], [68]. This virtual setting allows clinicians to run in silico tests of different treatment strategies, including choosing the right drug, adjusting dosage, and optimizing the timing of administration before applying them in real clinical practice. Advancing personalized therapies, especially for chronic and costly conditions, produce more effective and safer treatments [69]–[70]. This is highly valuable and appreciated. Digital twins model the complex dynamics of hospitals, such as patient flow, bed occupancy, staff allocation, and equipment use. This gives administrators a safe "sandbox" environment to test out changes to the logistics. The Mater Private Hospital in Dublin is a good example of how digital twin simulations were used to improve the workflows in the imaging department, which led to shorter wait times for patients and better use of equipment [71]–[72].

In oncology, digital cancer twins apply mathematical models of tumor growth and predictive analytics to estimate how a patient might respond to chemotherapy or radiotherapy[73].

With positive outcome simulations, clinicians are able to create individualized treatment protocols, reduce negative side effects from treatments, and improve the likelihood of positive treatment outcomes. This ultimately leads to more evidence-based outcomes[74].

On a population scale, digital health twins integrate diverse datasets like imaging and demographic data with machine learning to predict disease risk and advise on prevention. This includes information from genomics, lifestyle, and epidemiology. These kinds of platforms get patients more involved and help with resource allocation, which makes the whole system work better. In conclusion, digital health twins are essential to precision medicine and smart health management, as they deal with both individual and population health [75]–[76].

6. Neuroscience

Digital twins contribute to a new wave of personalized virtual brain projects in neuroscience that combine large-scale neural dynamics modelling with neuroimaging, genomics and clinical data .These systems provide an innovative way of studying the brain and tackling brain-related illnesses [34], [77]. In the EU, VirtualBrainCloud offers cloud computing services combining whole-brain models and population datasets and provides The Virtual Brain model. The integration of these systems allows for the personalized prevention and exclusive treatment of neurodegenerative illnesses, especially dementia, all in compliance with privacy and GDPR Regulation regulations [78]–[79]. Recent research in computational neuroscience involving digital twins derived from primate electrocorticography (ECoG) data has shown that hierarchical generative models, specifically variational recurrent neural networks (V-RNNs), can directly reconstruct transitions between states of consciousness, such as wakefulness and anesthesia, from neural recordings . These studies also demonstrate that virtual environments can facilitate the testing of pharmacological interventions or neuromodulation strategies in a safe and reproducible manner [13].

This study demonstrates the potential of digital brain models to support simulation-ready BCIs, predictive neurorehabilitation, and low-risk evaluation of brain interventions through multiscale integration of experimental and therapeutic data. In the field of BCIs, this helps with the design of neural prostheses, personalized rehabilitation plans, and the discovery of new neural biomarkers through virtual experimentation. This method is becoming a key part of precision neuro-medicine, intelligent rehabilitation, and the next generation of BCIs. It is based on dynamic, generalizable models that are continuously updated with real-world data (EEG, LFP, fMRI, and behavioral signals).

## E. Overview of Brain-Computer Interface Systems

A BCI system acts as an interface between the human brain and a computer system [1]. It measures brain activity and allows its users to control external devices that are independent of peripheral nerves and muscles with brain activity. Thus, the natural outputs of the central nervous system (CNS) are replaced, restored, enhanced, supplemented, or improved by an artificial output [80]–[81]. In general, at the beginning of the loop of a BCI system, there is a hardware including electrodes and a data acquisition system to transfer the received electrical activities in a digital and processable form to the computer system. After that, data processing steps and feature extraction will be applied, and then the extracted features are translated in such a way as to produce artificial output or commands that can create artificial output [82]–[83].

BCIs are generally classified into two main groups based on the method of neural signals acquisition: non-invasive and invasive systems [82], [84]–[85]. Recently, another category has emerged, known as hybrid BCI, which combine the two main methods to combine their advantages [86]–[87]. The general classification of BCIs and the recording technologies used in each of them is shown in Fig. 6.

### *1) Non-invasive BCIs*

Non-invasive BCIs are a class of BCIs that do not require surgery or invasive procedures in the data acquisition stages and operate only by recording neural activity from the surface of the head [88]. Despite the advantages such as ease of data acquisition and low cost, these methods usually have limitations in terms of spatial and temporal resolution; because the signals are attenuated when passing through different layers of the skin and skull. In general, non-invasive brain-computer interfaces use changes in the dynamics of brain oscillations such as event-related synchrony (ERD, ERS), steady-state evoked potentials (SSEPs), P300 evoked potentials and its related components, BOLD signals in real-time fMRI or oxyhemoglobin signals measured by near-infrared spectroscopy (NIRS) and other methods [88]–[90]. One of the most common methods is electroencephalography (EEG), which records the electrical potentials resulting from the activity of cortical neurons through surface electrodes. EEG is widely used in BCI systems due to its very high temporal resolution, low cost, and portability.

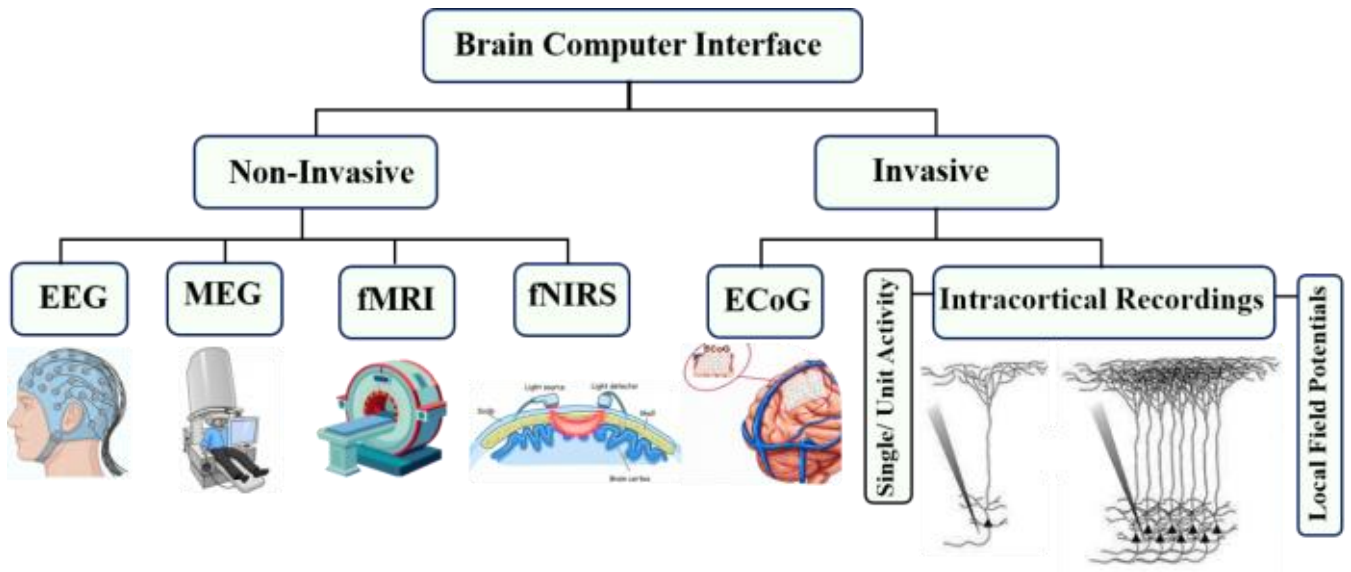

Fig. 6. Overview of invasive and non-invasive BCI categories based on signal acquisition methods.

However, its weakness in spatial resolution and high sensitivity to noise are its main limitations. The most important applications of EEG include P300 Speller systems for patient communication[91], Motor Imagery for controlling robotic limbs [92], SSVEP-based systems for rapid command selection [90], emotion recognition in Affective Computing [93], and neurorehabilitation.

Another non-invasive method is near-infrared spectroscopy (fNIRS), which indirectly estimates neural activity by measuring changes in hemoglobin concentration in cortical tissues. fNIRS is highly efficient in terms of electrical noise immunity and portability; however, its low temporal resolution is one of its most important challenges. It is commonly used in cognitive load assessment, attention monitoring, motor rehabilitation, and also in hybrid EEG–fNIRS systems to increase classification accuracy.

At a more advanced level, there is functional magnetic resonance imaging (fMRI), which measures the BOLD signal to examine activity in deep and superficial brain regions. Although fMRI provides very high spatial resolution, it is less widely used in practical BCI settings due to its low temporal resolution, high cost, and lack of portability. However, it can play an important role in brain mapping, decoding complex cognitive states, neurofeedback, and extracting prerequisites for building accurate models in digital brain twins. Finally, magnetoencephalography (MEG) records magnetic fields from neural activity, providing higher spatial and temporal resolution than EEG, while not suffering from cranial distortion. Its limitations include its high cost, large size, and the need for magnetically shielded rooms. MEG has wide applications in areas such as localization of activity sources, preoperative planning in epileptic patients, investigation of neural networks, and the study of brain oscillations.

### 2) *Invasive BCI*

Invasive BCIs allow for direct and highly accurate recording of neuronal activity by placing electrodes on the surface of the cerebral cortex or within neural tissue, using cortical arrays, ECoG, spinal cord implants, and the like[80], [94]–[95]. These methods are used in advanced research and sensitive clinical applications due to their good spatial and temporal resolution. However, their invasive nature brings with it surgical risks, long-term stability issues, and significant ethical considerations[94], [96]–[97]. One of the most widely used invasive recording methods is ECoG, which records by placing an electrode array on the surface of the cortex[98]. Due to its lower sensitivity to noise and higher resolution, this recording method is widely used in detecting epileptic seizure locations, predicting seizures, and decoding motor movements for controlling robotic limbs. ECoG data have also been used as one of the first sources for designing early digital brain twin models because they reflect cortical network dynamics with reasonable accuracy. For example, a recent study using ECoG data from macaque monkeys and an advanced model based on a digital brain twin enabled real-time simulation of cortical signals [13].

Another common invasive recording method is local field potentials (LFPs), which record the activity of populations of neurons in specific brain regions[99]. Because LFPs have a high signal-to-noise (SNR) ratio, they are used in decoding force and motor state, studying oscillatory mechanisms associated with movement, and understanding and developing deep brain stimulation (DBS) systems for disorders such as Parkinson's disease[100]. In addition, intracortical recordings (Spikes) allow the recording of action potentials of single neurons using microelectrode arrays such as the Utah array. This method provides the most precise level of access to neural activity and is therefore useful for applications such as precise control of robotic limbs, restoration of communication and movement in severely paralyzed patients, and tuning neural models in digital twin simulators[94], [101]. However, this method is the most invasive recording method and poses challenges such as surgical risk and long-term stability of the electrodes [96], [106]–[107].

### 3) *Hybrid BCI*

In recent years, research in the field of BCIs has mainly moved towards the development of hybrid BCIs. In hybrid BCIs, the main approach is to simultaneously integrate multiple neural recording methods to overcome the limitations of single recording methods and provide a more accurate picture of brain activity[83]. For example, the combination of EEG and fNIRS is one of the most common examples which improves the decoding performance of the motor imagery of both force and speed of hand clenching by combining the high temporal resolution of EEG and the hemodynamic information of fNIRS [102]–[103].

Furthermore, the combination of EEG recording with fMRI imaging, by combining high temporal resolution in EEG and good spatial resolution in fMRI, provides a precise spatiotemporal map of functional and structural information of the brain [104]–[105]. Such a combination is in a special position for decoding cognitive states and creating digital twins, since fMRI will act as a provider of anatomical information and EEG as a source of dynamic signals.

The integration of invasive recording methods will enable the analysis of cortical and subcortical structures and will lead to the advancement of controlled neuroprosthetic systems. This class of hybrid BCIs, which are built by combining methods including EEG and fMRI are becoming increasingly important, especially for medical and research systems that require high accuracy and real-time response.

# IV. Methododology

## A. Major Challenges of BCI Systems

BCI systems offer substantial benefits; however, they confront fundamental limitations that hinder their scalability and real-world deployment. Some important limits are:

*1)* **Real-time processing latency:** The computational overhead of signal preprocessing, combined with the inference time of deep architectures like CNNs and LSTMs, inevitably introduces system latency. It is precisely these millisecond-level delays that threaten to destabilize the closed-loop feedback essential for seamless control. We are, therefore, forced into a rigid trade-off between decoding precision and speed—a delicate balance where even marginal lags can sever the user's sense of agency[80], [106]–[110].

*2)* **High surgical risks and invasiveness:** The placement of electrodes for invasive BCIs necessitates surgery and the associated medial risks and medical costs including infection, bleeding, and other medical complications [96]–[97], [111]–[112].

*3)* **Hardware and biocompatibility constraints:** The major delivery aspect in terms of invasive BCIs is the recording interface. This relates to the other major issues of limited or no biocompatibility, electronic disulfides or other materials ‘atrophy’ or soften Murphy interacting tissues, and other electronic system malfunction or breakdown mas [97], [113]–[116].

*4)* **Calibration instability and individual variability:** A considerable amount of recalibration and tuning is needed in BCI to counter drift and maintain system performance, however, inter subject differences in such adaptive drift during implant recording and interfacing systems is such a considerable challenge [110], [117]–[119].

*5)* **Lack of standardization:** The field continues to be impacted by issues surrounding the heterogeneity of approaches to system design. An inability to compare and validate multiple BCI systems due to the wide disparity in feature extraction strategies and performance metrics is a major bottleneck in the field [84], [97], [120]–[121].

## B. The Digital Twin Paradigm: A Transformative Solution

To overcome these limitations, recent research has begun to integrate BCI technology with DT paradigm, aiming to create virtual and dynamic copies of biological systems.Within neuroscience, the Digital Twin-based BCI (DT-BCI) focus on the inner neural dynamics, and system interaction with external devices simultaneously. Within neuroscience, DT-BCI focus on the inner neural dynamics, and system interaction with external devices simultaneously.This approach offers several distinct advantages:

*1)* Continuous adaptation: It allows for the continuous optimization of the two-way performance between the brain and the machine.

*2)* Personalized modeling: Individual neural modeling for every person is made possible.

*3)* Predictive simulation: It enables the simulation of cortical activity under various task conditions and allows for the prediction of signal changes or electrode degradation before the system's performance is impaired.

*4)* Multimodal integration: Systems that employ the DT framework robustly integrate multimodal data (e.g., EEG and fMRI behavioral data). This improves system interpretability, and makes them more acceptable on ethical grounds.

As can be seen in Fig. 7, the fundamental distinction between the two approaches lies in the system’s capacity for dynamic adaptation. The conventional framework (Fig. 7(a)) typically operates as a linear, static pipeline where decoding parameters remain fixed post-training, limiting the system's ability to respond to evolving neural states. In contrast, the proposed BCI-DT framework (Fig. 7(b)) introduces a paradigm of continuous adaptivity via the Digital Thread. This bidirectional link establishes a symbiotic loop between the real and digital brains. Unlike the rigid structure of conventional models, this architecture utilizes the Digital Brain to run parallel simulations, allowing the system to dynamically recalibrate and optimize its decoding strategy in real-time, independent of immediate input fluctuations.

Thus, intersection of BCI and DT is a promising innovation, moving systems to flexible, self-learning, and highly robust clinically systems for brain machine interfacing.

The subsequent sections provide a rigorous formalization of the Digital Twin concept, followed by a critical exposition of frameworks specifically calibrated for BCI applications, with a particular emphasis on neuroscientific axioms. We then delineate the architectural topology of neuro-DT systems, concluding with a comprehensive analysis of applications that synergize digital twin methodologies with advanced neuroscience.

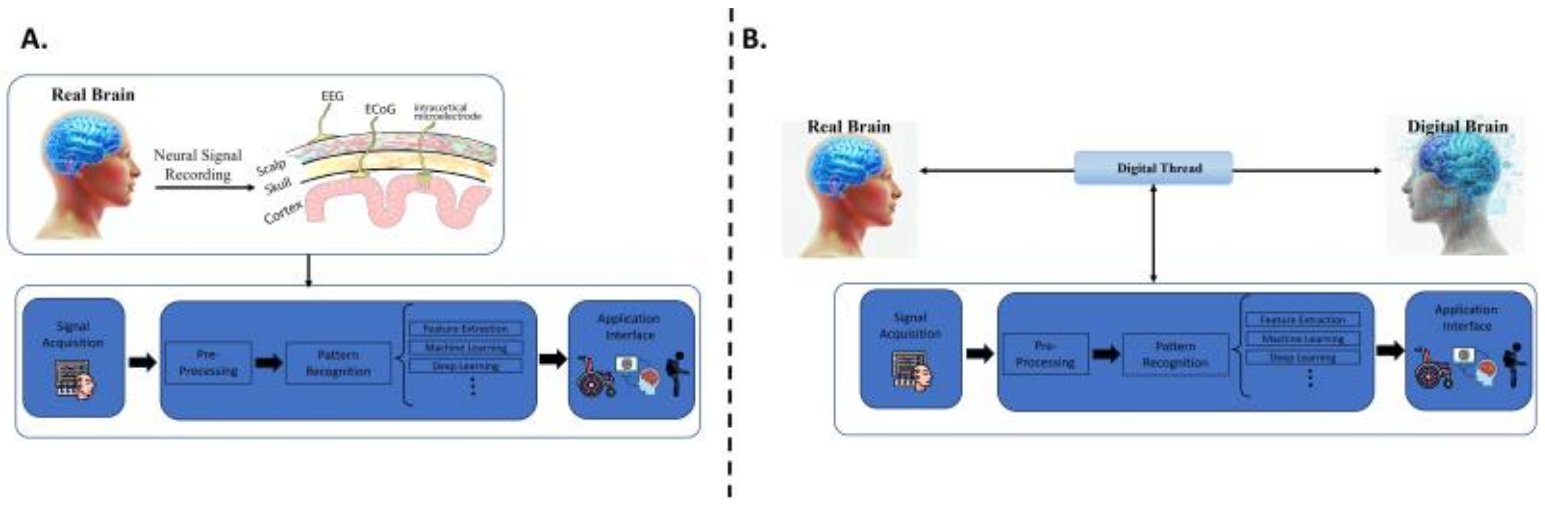

Fig. 7. Comparison between conventional BCI and BCI with digital twin (BCI-DT). (a) Conventional BCI. (b) Proposed BCI-DT.

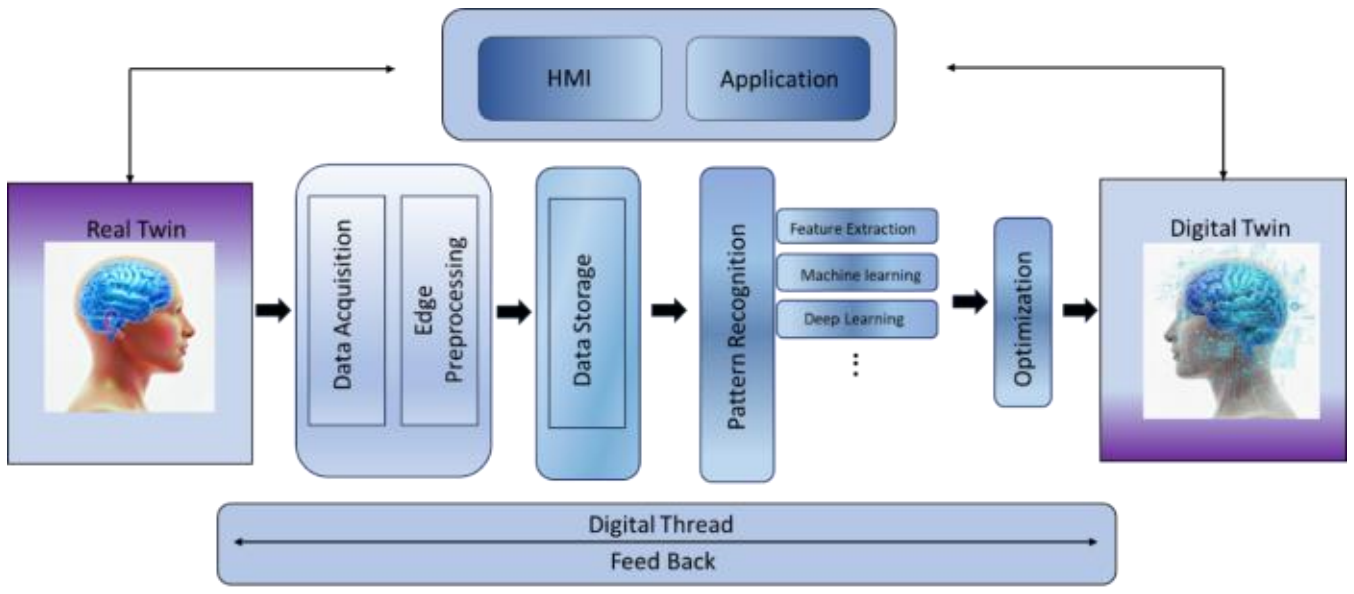

Fig 8. Structure of the proposed method, BCI-DT. The figure illustrates the layered architecture of the model.

## C. Proposed Method: BCI-DT

Both BCI and Computational Neuroscience fields are major sub-branches of Neuroscience. Therefore, to create a comprehensive and integrated view, we must first consider the interaction between Neuroscience and the emerging concept of DT. In this regard, we present a framework called Digital Twin for Neuroscience, which we have briefly introduced as Neural Digital Twin (NDT). In fact, NDT can be considered the evolved generation of the field of computational neuroscience; a system that goes beyond mere modeling or simulation and is known as a dynamic, intelligent, and synchronized digital representation of the brain and neural activities. By leveraging this framework, a new generation of brain-computer interfaces can be developed, known as BCI-DT (Brain–Computer Interface based on Digital Twin).

In this new generation, the dynamic and continuous interaction between the brain, the digital model, and the computational system enables continuous learning, personalized adaptation, and intelligent real-time feedback [132]. From this perspective, BCI-DT can be considered a fundamental step toward the evolution of brain-computer interfaces, as it has the potential to address many of the challenges of classic BCIs, including instability, high noise, and lack of individualization, by relying on the Neural Digital Twin architecture .

This system is built upon five complementary and interconnected layers: three main layers including the Physical Layer, the Virtual Layer, and the Communication Layer and two supplementary layers, namely the Analysis Layer and the Feedback Layer. Together, these five layers form a closed-loop system that enables live, two-way, and continuous interaction between the real world (the brain and neural signals) and its digital representation. The structure of the proposed method, BCI-DT, is shown in Fig. 8.

In the following, each layer will be described in detail.

*1) Physical Entity*

In a neural digital twin, this layer is the foundation and main source of data and includes the brain, nervous system, and neural signal recording devices, such as ECoG, LFP, or EEG electrodes, motion sensors, and neural prostheses. The main function of this layer is to generate live data that directly reflects neural activity and physical body function. The characteristics of this layer are as follows:

- High sensitivity and precise temporal and spatial resolution for recording neural signals.
- Ability to execute feedback commands sent from the feedback layer.
- Stability and connectivity to digital networks to ensure uninterrupted data flow.

In addition, the data generated by this layer is transmitted to the virtual and analytical layers through a kind of digital thread, and corrective feedback is returned to it from the feedback layer.

*2) Virtual Representation*

This layer creates a dynamic digital representation synchronized with the real system, going beyond mere modeling and simulation. Its main function is to synchronize real-world data with the digital model and provide a real-time snapshot of neural activity. Key features of this layer include:

- Real-time synchronization with physical layer data.
- High flexibility to incorporate scientific models, deep neural networks, and AI-based prediction algorithms.
- Scalability for interactive holograms and shadow simulations to observe and analyze system behavior.

In this layer, data is received from the physical layer. Then, the received data is prepared for analysis at this stage, and the results of the analysis and predictions are transmitted to the feedback layer. Also, in this layer, real-time changes and signals are managed bidirectionally through the digital thread.

*3) Digital Thread*

The digital thread is a live, two-way communication channel between the physical and virtual worlds. Its main function is to transmit live data, commands, and feedback in real time, ensuring that the system operates in a closed loop. The specifications and features include:

- Secure, low-latency data transmission, which is essential for real-time interaction.
- Data integration and standardization to ensure consistency between the physical and virtual layers.
- Network redundancy and stability to prevent interruptions in the feedback loop.

The digital thread simultaneously transmits data from the physical layer to the virtual and analytical layers, and returns the feedback generated in the feedback layer to the physical layer.

*4) Analysis Layer*

In this layer, the collected data is analyzed and then the corresponding neural patterns are identified. In fact, in this layer, the main task is to identify the functional states of the brain, motor intention patterns and other neural indicators that provide detailed insights for system optimization. The characteristics of this layer are as follows:

- Use of deep learning algorithms, graph neural networks and advanced statistical analysis methods.
- Ability to detect and predict neural behaviors and intentions in real time.
- Provide interpretable outputs and use to improve BCI performance.

Input data reaches this layer through a digital thread and then the analysis and results from this layer are sent to the feedback layer for corrective actions. This layer can also share the results to improve the virtual model.

*5) Feedback Layer*

The feedback layer is responsible for returning optimal and corrective decisions to physical reality and completing the closed loop. This layer enables adaptive and intelligent control, allowing BCI-DT to consider individual performance and environmental conditions. Features and Specifications:

- Real-time feedback to physical sensors and actuators.
- Adaptability to environmental changes and individual user characteristics.
- Ability to integrate with AI-based decision-making algorithms for performance optimization.

Feedback generated from the analysis layer is taken from it and returned to the physical layer via the digital thread. Feedback information can also be used to improve and synchronize the virtual layer.

### D. Summary of Layer Communications and Functions

These five layers in NDT act as a dynamic, two-way ecosystem: real data is collected from the physical layer and transmitted through the digital thread to the virtual and analysis layer, analyses and predictions are sent to the feedback layer and applied to the physical system, and the results of these applications enter the loop as new data. This continuous interaction and precise synchronization form the basis of the new generation of BCIs, namely BCI-DTs, and allow overcoming the challenges of classical BCIs.

Thus, this five-layer architecture operates as a closed, dynamic loop. In this loop, neural data continuously flows from the physical environment and is analyzed at each stage, while the results of processing and analysis are also fed back to the same environment, changing the real state of the system. This process begins from the moment the signals are captured live and continues by transforming this data into a dynamic digital representation that constantly interacts with the real world. This representation is not a static image, but an evolutionary model that reflects environmental changes and at the same time provides the necessary information for intelligent analysis. The analysis layer uses this platform to extract effective features, predict neural dynamics and make decisions in the moment; decisions that are then applied as corrective feedback to the real environment, leading to new changes in the signals. As a result, the connection between the physical world and the digital representation is a fully two-way and continuous relationship that keeps the system in dynamic equilibrium.

The role of the communication layer in this context is time coordination, secure transmission and data integrity. Thus, the system with a feedback mechanism will be able to operate seamlessly with minimal delay. Such a structure forms the basis of an intelligent and self-adaptive neural ecosystem capable of continuous learning, behavioral prediction, and instantaneous response. Thus, such a system can ultimately overcome the limitations of traditional BCIs in areas such as stability, accuracy, and processing efficiency.

## V. Application

In conclusion, it can be inferred that Digital Twin, as a dynamic and personalized framework that blends real-time data streams with the power of AI-driven prediction and simulation, has the potential to effectively address the fundamental challenges previously outlined for BCI systems. This integration creates not just a digital representation, but a responsive and bidirectional system that continuously learns, corrects, and predicts neurological states.

Integrating DT with BCI can revolutionize how neural decoding is performed, as this architecture allows for the creation of models specifically adapted to each individual, reduces noise factors and biological variability, learns individual neural patterns over time, and optimizes rehabilitation and neural control processes. Table 1 summarizes the main aspects, which are further discussed in the next section.

### A. Brain-Computer Interface Integration

The integration of generative brain models into neurotechnology frameworks has gained attention for enhancing the adaptability of closed-loop BCIs. The Virtual Brain Twin (VBT) platform proposed by Hashemi et al. bridges mechanistic modeling and data-driven inference, enabling rapid personalization and in silico optimization of decoding and stimulation strategies prior to clinical deployment [34]. Its capacity to simulate brain dynamics under diverse perturbations further supports causal analysis while reducing reliance on extensive biological experimentation.

This capability is illustrated by recent neural digital twin initiatives, including the Living Brain Project, which explores real-time large-scale neural simulations for adaptive brain–machine interaction. Such platforms demonstrate the potential of predictive modeling to support optimization of stimulation strategies [129]. Similarly, the Digital Twin Brain Simulator leverages primate ECoG data to study brain-state transitions relevant to adaptive intracortical BCIs [13].

Fundamentally, these systems engender a bidirectional feedback loop: the physical BCI transmits multimodal data into the NDT, which reciprocates with predictive insights to refine decoding algorithms. This synergy has been shown to substantially reduce calibration time while enhancing the long-term stability required for robust neuroprosthetic control, including applications targeting locked-in syndrome.

### B. Personalized Brain Modeling for Precision Medicine

The convergence of NDTs in healthcare has accelerated the shift from population-based heuristics to precision medicine, spanning early diagnosis to therapeutic intervention. For monitoring and early detection, the Digital Twins for Alzheimer's initiative (Harvard Medical School & MGH) emphasizes digital phenotypes by integrating speech and cognitive data [130]–[131], supporting predictive models capable of identifying subtle cognitive decline before overt symptoms. In pharmacology, the Digital Twins for Alzheimer's Drug Discovery project (Cleveland Clinic) leverages mechanism-based virtual trials to account for biological heterogeneity, enabling virtual drug screening and biomarker exploration to guide treatment strategies [132]. At the intervention level, the Virtual Epileptic Patient (VEP, EBRAINS) combines patient-specific MRI data with Bayesian and simulation-based inference to identify epileptogenic zones and test virtual surgical procedures [129], while the Neurotwin Project (EU Consortium) applies laminar-resolution whole-brain models to tailor non-invasive interventions, predicting patient responses to tES/TMS prior to clinical trials [137].

### C. Predictive Modeling of Neurological Disorders

Predictive modeling constitutes a cornerstone of NDT utility, facilitating the in silico emulation of disease trajectories and therapeutic responses well before clinical manifestation. A prime example is the 'Virtual Brain Twins for Stimulation in Epilepsy' (EBRAINS; 2024), which achieves high-precision localization of epileptogenic foci through patient-specific simulations [129]. By enabling the virtual testing of stimulation paradigms, this framework has demonstrated the capacity to substantially mitigate surgical risks.

In the domain of neurodegeneration, Stanford's digital twin initiatives (USA; 2023–2027) elucidate hyper-excitability patterns in learning disabilities and cortical dynamics in preclinical Alzheimer's models, effectively forecasting neurocognitive deterioration years in advance. Complementing this, the 'Brain Bio-Digital Twin' (NTT Corporation and NCNP, Japan; 2020–ongoing) employs advanced machine learning on multimodal data to detect prodromal signs of dementia and depression, thereby operationalizing platforms for robust risk stratification [136].

These predictive capabilities extend to acute clinical events, as evidenced by the 'Digital Twin Brain Simulator' (RIKEN-EBRAINS; 2025), which provides real-time consciousness monitoring to accurately model anesthesia transitions and inform critical-care decision-making [13]. Collectively, these models catalyze a fundamental shift in neurology from reactive symptom management to proactive, preemptive prevention.

### D. Neuroscience Research and Hypothesis Testing

NDTs serve as powerful virtual laboratories, enabling the exploration of mechanistic hypotheses that are often infeasible under biological constraints. The mouse visual cortex digital twin [127] represents a compelling example for basic neuroscience research and hypothesis testing, as it can predict neural population responses to novel stimuli and support rapid, large-scale in silico experimentation while reducing reliance on extensive in vivo recordings. Through virtual perturbations and parallel evaluation of diverse stimulus conditions, such frameworks allow efficient validation of hypotheses related to neural encoding, circuit organization, and functional connectivity, thereby accelerating discovery and minimizing animal experimentation.

Similarly, the University of Cambridge's 'BrainTwin Project' (2023–2026) models the complexities of sleep and psychiatric conditions, allowing for the *in silico* testing of causal links in mood regulation [134]–[135]. By establishing a reproducible and fully controllable experimental environment, these systems drastically accelerate the discovery process—reducing the trajectory from hypothesis generation to validation—while simultaneously reducing the need for animal experimentation. This approach is revolutionizing systems neuroscience, as seen in initiatives like the 'Living Brain Project,' which permit the comprehensive analysis of emergent properties such as consciousness and decision-making [129].

### E. Future Directions

NDTs have the potential to enable adaptive human–machine interfaces that extend beyond therapeutic applications to real-time, personalized neurotechnologies. Near-term priorities include improving model fidelity toward million-neuron-scale resolutions and exploring neuromorphic computing for efficient real-time emulation [138]. Federated learning frameworks are expected to play a key role in preserving data privacy across large, multinational datasets.

The European Virtual Human Twins (VHT) Initiative exemplifies this vision, developing multiscale digital representations of human physiology that include predictive neural models for personalized interventions. Within this framework, projects such as the Virtual Brain Twin platform generate patient-specific brain models incorporating neuroplasticity and aging effects, supporting closed-loop BCIs [34]. Hybrid systems that integrate invasive high-bandwidth implants with non-invasive neuromodulation techniques may leverage population-level NDTs to optimize virtual clinical trials and therapeutic protocols [34].

As NDTs mature, they could serve as platforms for cognitive augmentation, highlighting the need for rigorous ethical oversight concerning personal identity, autonomy, and equity. Interdisciplinary governance mechanisms will be essential to ensure equitable access and mitigate potential risks. Overall, NDTs illustrate a convergence of neuroscience and advanced computing, redefining the boundaries of human–machine interaction.

## VI. Conclusion

BCI system is an interface between the human brain and a computer system that measures brain activity and allows the user to control external devices independently of their peripheral nerves and muscles and solely with brain activity to replace, restore, augment, supplement, or improve the natural outputs of the CNS with an artificial output[1]. These systems have many fundamental limitations, including lack of real-time processing, hardware limitations, and surgical risks for electrode implantation, which reduce their long-term usability, especially in clinical and neurorehabilitation settings[122]–[123]. The concept of a NDT offers an innovative approach to overcome the challenges of traditional BCIs.

By creating a dynamic and personalized computational model that can be updated according to brain changes, real-time monitoring, adaptive decoding of neural activity, prediction of brain state, and intelligent feedback in the brain digital twin are possible. NDTs consist of a physical layer, a virtual layer, a digital thread, an analysis layer, and a feedback layer, which interact with each other. In this framework, neural data is continuously collected and analyzed, and after corrective actions are taken, they apply changes to the physical environment through a feedback mechanism, resulting in synchronization between the real and digital worlds. The neural digital twin is a fundamental evolution in the design and implementation of BCI systems. By creating continuous interaction and two-way feedback between the real and digital worlds, NDT-based systems can overcome the limitations of traditional BCIs and pave the way for the integration of neurotechnology in medicine, rehabilitation, and advanced brain-machine applications[13], [75], [124], [125].

This approach promises a future in which brain-computer interfaces with higher accuracy, stability, and accessibility will become a new generation of neural control, predictive modeling, and self-adaptive neurostatic systems [75], [126].

| Project Name | Lead Institution & Country | Brief Description | Key Outcomes |
|---|---|---|---|
| **Stanford Digital Twin of Mouse Visual Cortex**[127] | Stanford University (USA) | AI-driven digital twin of mouse visual cortex replicating neural activity with near-physiological precision for visual processing and BCI insights | Foundational for scalable mammalian brain twins; key insights into neural encoding, 2025 |
| **Stanford Digital Twins for Math Disabilities**[128] | Stanford Cognitive & Systems Neuroscience Lab (USA) | Personalized brain digital twins in children to identify neural mechanisms of mathematical learning disabilities | Clinical translation for educational interventions, 2024 |
| **Living Brain Project**[129] | Dassault Systèmes, MIT, NVIDIA (France/USA) | Real-time digital twin of human brain with 100,000 simulated neurons targeting BCI and neurological disorders | Phase 1 results, demonstrate utility in dementia modeling and neuromodulation, 2025. |
| **Digital Twins for Alzheimer's**[130]–[131] | Harvard Medical School & MGH (USA) | Speech- and cognition-based digital twins for Alzheimer's patients enabling predictive chatbots and disease progression modeling | Ongoing clinical validation; featured in institutional reports, 2025 |
| **Digital Twins for Alzheimer's Drug Discovery**[132] | Cleveland Clinic & Case Western Reserve University (Feixiong Cheng Lab) / USA | Utilization of precision medicine digital twins for virtual drug screening, biomarker discovery, and repurposing in Alzheimer's disease | Review of DT applications in AD drug development; highlights reduced trial failure rates and personalized therapy, 2024 |
| **Virtual Brain Twin (Psychosis)**[34], [124], [133] | EBRAINS AISBL & Aix-Marseille Univ. (Europe) | Virtual brain twins for personalized therapy in schizophrenia and psychosis using AI-driven microcircuit simulation | €10 M funding; published in *Nature Computational Science,* 2025 |
| **Virtual Brain Twins (Epilepsy)**[129] | EBRAINS & Partners (Europe) | Patient-specific brain twins for non-invasive prediction of epileptogenic zones and stimulation therapy | High-accuracy EZN localization; published in *Nature Computational Science,* 2025 |
| **Twin4Health**[70], [76] | Karolinska Institutet (Sweden) | Hospital-wide digital twin with emphasis on brain and cardiac modules – first fully operational clinical deployment | Entered clinical phase; resource optimization reported, 2025 |
| **BrainTwin Project**[134]–[135] | University of Cambridge (UK) | Digital twin approach for sleep disorders and psychiatric conditions | Integrated into Accelerate Science program; ongoing modeling |
| **Digital Twin Brain Simulator (Primate ECoG)**[13] | RIKEN & EBRAINS (Japan/Europe) | Real-time consciousness monitoring and virtual intervention simulator using primate electrocorticogram data | High-fidelity anesthesia transition modeling; published in *npj Digital Medicine,* 2025 |
| **Brain Bio-Digital Twin** [136] | NTT Corporation & NCNP (Japan) | Bio-digital twin of the brain using AI/ML for early detection of dementia, depression, and neurological diseases | Operationalized platform; established early-stage validation for personalized diagnostic stratifications, 2024 |
| **Neurotwin Project** [137] | Neuroelectrics & EU Consortium (Spain/EU) | Optimizing non-invasive stimulation for Alzheimer's using laminar-resolution whole-brain models | Validated personalized progression models; advanced protocols to clinical trial phase for therapeutic efficacy, 2024 |

*Table 1. Landscape of State-of-the-Art Neural Digital Twin Ecosystems: Towards Next-Generation Brain-Computer Interfaces and Precision Medicine.*

## ACKNOWLEDGMENT

Not applicable.